\documentclass[pre,twocolumn,floatfix,showpacs,superscriptaddress,aps]{revtex4}
\usepackage{amsmath,graphicx}
\usepackage{color}

\begin{document}

\title{Multistable behavior above synchronization in a locally
coupled Kuramoto model}
\author{Paulo F. C. Tilles}
\affiliation{Instituto de F\'{i}sica Te\'{o}rica UNESP - Universidade Estadual Paulista,Caixa Postal 70532-2,
01156-970 S\~{a}o Paulo, SP, Brazil}
\author{Fernando F. Ferreira}
\affiliation{Grupo Interdisciplinar de F\'{\i}sica da
Informa\c{c}\~ao e Economia
 GRIFE, Escola de Arte, Ci\^encias e Humanidades,
 Universidade de S\~ao Paulo,  Av. Arlindo Bettio 1000, 03828-000 S\~ao
 Paulo, Brazil}
\author{Hilda A. Cerdeira}
\affiliation{Instituto de F\'{i}sica Te\'{o}rica UNESP - Universidade Estadual Paulista,Caixa Postal 70532-2,
01156-970 S\~{a}o Paulo, SP, Brazil}

\begin{abstract}
A system of nearest neighbors Kuramoto-like coupled oscillators placed in a ring is studied above the critical synchronization transition. We find a richness of solutions when the coupling increases, which exists only within a solvability region (SR). We also find that they posses different characteristics, depending on the section of the boundary of the SR where the solutions appear. We study the birth of these solutions and how they evolve when {K} increases, and determine the diagram of solutions in phase space.
\end{abstract}

\maketitle

\section{Introduction}

The ubiquity of phenomena linked to coupled chaotic systems has made them the focus of interest for the last twenty years. The study of these systems has raised interest with the intent of realistically modeling spatially extended systems, as diverse as Josephson junction arrays, multimode lasers, vortex dynamics, biological information processing, neurodynamics as well as applications in communications \cite{sakag,erment,daido,winfree,cwwu,shstrog,manru,arenas}, with the belief that dominant features will be retained in such simple models. Coupled systems with local interactions are of special importance. In particular the Kuramoto model\cite{kuramoto} in its local version: locally coupled Kuramoto model (LCKM) has raised attention since most features of systems with phase coupling appear in this particularly simple model \cite{kurths,hugang,zheng,liu,acerbon}.

This system which shows synchronization in the mean frequency has
been thoroughly studied at and before full synchronization. The oscillators
cluster in average frequency decreasing the number of clusters until they
come into a single cluster at full synchronization. At this moment the frequency becomes constant
and the phases lock, such that all phase differences are constant. The
solution for full synchronization and its stability has been studied by many
authors \cite{strog01,daniels,praman,hassan02,hassan01}. Zheng et al \cite{zheng} already in 1998 pointed out that the
behavior of the order parameter "indicates the coexistence of multiple
attractors of phase locking states" above the synchronization critical
coupling, $K_{s}$. Synchronization with coexistence of attractors has been
reported by different authors and in different fields below $K_{s}$ \cite
{bocal1,hugang,maurer,heff,vibert,perli,freudel,liu,zheng}
but, to the best of our knowledge nobody else has pursued the matter above
full synchronization. If we have the intention to simulate real systems,
mostly in technological applications, it is important to know whether or not
we can move freely below and also above synchronization within stable
solutions and to know whether or not they are unique. In this work we shall
study a LCKM above complete synchronization, we shall show that there is an
unexpected richness of behavior: multistable solutions appear and we cannot
change the strength of the coupling without danger of falling into different
attractors.

\section{Locally coupled Kuramoto model on the synchronized region}

The model that we use is described by the following equations:
\begin{equation}
\dot{\theta}_{n}=\omega _{n}+K\left[ \sin \left( \theta _{n-1}-\theta
_{n}\right) +\sin \left( \theta _{n+1}-\theta _{n}\right) \right] ,
\label{1}
\end{equation}%
where \ $n=1,...,N$, and $\omega _{n}\in \left\{ \omega \right\} $ is the
set of natural frequencies of the oscillators. The ring topology is defined
by the periodic conditions $\theta _{N+1}=\theta _{1}$\ and $\theta
_{0}=\theta _{N}$. There is a minimum value for the coupling constant $K$,
denoted as critical synchronization coupling $K_{s}$, that drives the system
into a fully synchronized state \cite{praman,hassan01,hassan02}. In this state the oscillators instantaneous frequencies assume a constant
value $\Omega =\frac{1}{N}\sum_{j=1}^{N}\omega _{j}$\ that remains unchanged
for any $K\in \left[ K_{s},\infty \right) $.

The set of equations (\ref{1}) in a synchronized state may be written as
\begin{equation}
\frac{\Omega -\omega _{n}}{K}=\sin \phi _{n-1}-\sin \phi _{n}.  \label{2}
\end{equation}%
with $\phi _{n}=\theta _{n}-\theta _{n+1}$ and\ $n=1,...,N-1$ that enables
one to write the phase locking conditions as $\phi _{n}=\phi _{n}\left(
K,\left\{ \omega \right\} \right) $, with multiple stable solutions
depending both on the number of oscillators $N$ and on the coupling constant
$K\geq K_{s}$. On a synchronized state any $\phi _{n}$\ may be written as a
function of an arbitrarily chosen distance $\phi _{n^{\ast }}$:
\begin{subequations}
\label{3}
\begin{equation}
\phi _{n}=\arcsin \left[ \sin \phi _{n^{\ast }}+\frac{1}{K}%
\sum_{j=n+1}^{n^{\ast }}\left( \Omega -\omega _{j}\right) \right] ,
\label{3a}
\end{equation}%
for $n=1,...,n^{\ast }-1\ $\ and
\begin{equation}
\phi _{n}=\arcsin \left[ \sin \phi _{n^{\ast }}-\frac{1}{K}\sum_{j=n^{\ast
}+1}^{n}\left( \Omega -\omega _{j}\right) \right] ,  \label{3b}
\end{equation}%
for $n=n^{\ast }+1,...,N-1$. With the identity $\sum_{j=1}^{N}\phi _{j}=0$\
it is possible to write $\phi _{N}$\ as the sum of all others $\phi _{n}$,
and the structure of the system allows us to reduce the set of $N$ equations
(\ref{2}) to a single equation on two variables $\left( \phi _{n^{\ast
}},K\right) $:
\end{subequations}
\begin{equation}
\sin \left( \phi _{n^{\ast }}+\sum_{n\neq n^{\ast }}^{N-1}\phi _{n}\right)
+\sin \phi _{n^{\ast }}=\frac{\sum_{j=1}^{n^{\ast }}\left(\omega_{j} - \Omega \right) }{K}.  \label{4}
\end{equation}%
The choice of $\phi _{n^{\ast }}$ will become apparent in the next paragraph.

If one removes a single link (interaction) between any pair of oscillators the result is a chain with free ends. If we can find the chain encapsulated inside the ring with the highest synchronization coupling we may be able to find the solutions above $K_{s}$. For each of the $N$ possible ways for which this procedure may be done, we label the oscillators as $\omega_{1},...,\omega _{N}$, keeping their order when we change the labels. Following the calculation of Strogatz and Mirollo \cite{strog01} we find the coupling constant at complete synchronization for a given chain as $K_{s}(l)=\max_{1\leq l\leq N}\left\vert \sum_{j=1}^{l}\left( \Omega -\omega_{j}\right) \right\vert $. Therefore we identify $n^{\ast }$ as the value of $l$ when this expression is the highest among all the synchronization couplings $K_{s}(l)$ and we call it $K_{s}^{\text{chain}}$.

\begin{figure}[!ht]
\centering\includegraphics[width=1.0\linewidth]{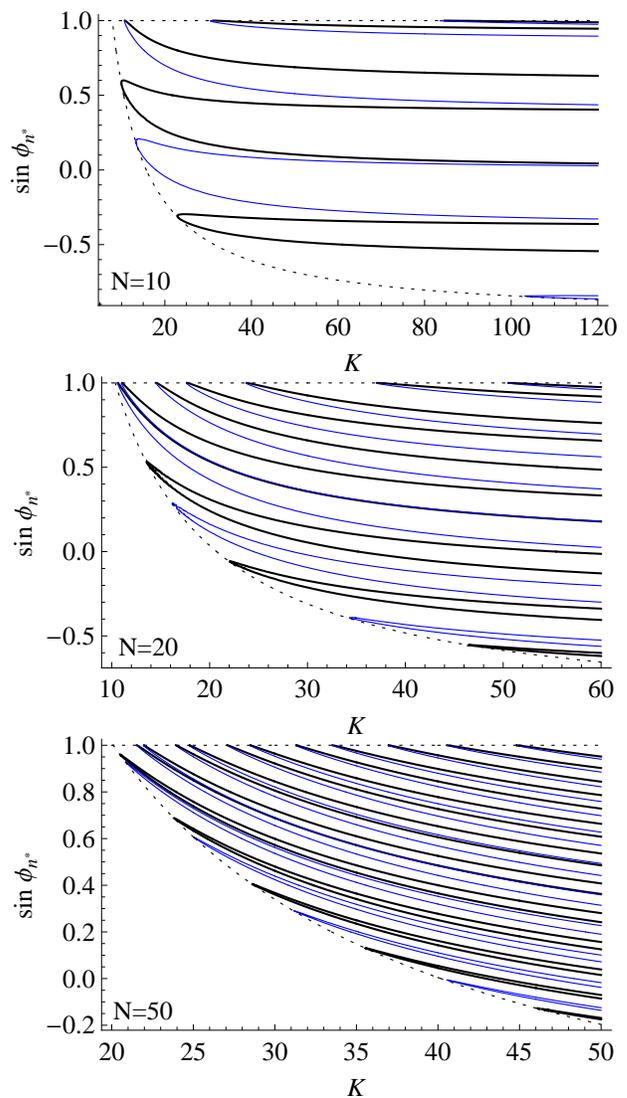}
\caption{(Color online) Numerical solutions of (\ref{4}) representing the bifurcation diagram for a ring of oscillators (stability not explicit): Top: $N=10$ with natural frequencies $\omega _{1}=6.9$, $\omega _{2}=2.8$, $\omega _{3}=-0.4$, $\omega _{4}=-2.6$,
$\omega _{5}=1.3$, $\omega _{6}=-6.8$, $\omega _{7}=0.8$, $\omega _{8}=-1.6$%
, $\omega _{9}=-9.5$ and $\omega _{10}=-6.7$. Middle: realization with $N=20$. Bottom: realization with $N=50$. The sets of natural frequencies $\left\{ \omega \right\} _{N}$ are generated from a uniform distribution function defined in the interval $\left[ -10,10\right]$. Black lines represent solutions with $\cos \phi _{n^{\ast }}>0$, blue lines are have $\cos \phi _{n^{\ast }}<0$ and the dotted lines are the limiting boundaries $\sin \phi_{N}=-s$ and $\sin{\phi_{n^{\ast }}}=s$.}
\label{figA}
\end{figure}

We write equation (\ref{4}) as follows:
\begin{equation}
\sin \phi _{n^{\ast }}-\sin \left[ \phi _{N}\left( \phi _{n^{\ast }}\right) %
\right] =s\frac{K_{s}^{\text{chain}}}{K},  \label{5}
\end{equation}%
where $s=sign\left[ \sum_{j=1}^{n^{\ast }}\left( \Omega -\omega _{j}\right) %
\right] $, and solve it numerically using Mathematica. The general form of the solution is presented in figure \ref{figA} for $N=10,20$ and $50$ where the frequencies $\left\{ \omega \right\} _{N}$ were generated from a uniform distribution function defined in the interval $\left[ -10,10\right]$. We have performed calculations for different natural frequencies realizations, all the results present the same features as will be described, therefore we keep here to a single realization for a given $N$ to avoid confusion.

We see that there are multiple phase locking solutions for the system above $K_{s}$ being spontaneously generated on a confined region on the bifurcation diagram (BD), as indicated by the dotted lines. These solutions are of two kinds: a) we call type I solutions the ones generated on the bottom boundary of BD with $\sin \phi _{n^{\ast }}\neq s$ that bifurcate into branches keeping $sign\left( \cos \phi _{n^{\ast }}\right) $ invariant; b) solutions with opposite signs of $ \cos \phi _{n^{\ast }}$ sharing the same origin on the top of BD (close to $\sin{\phi_{n^{\ast }}}=s$) are called type II.

The confinement region may be obtained realizing that for each value of $\sin \phi _{n^{\ast }}$ there is a maximum value of $K$ below which (\ref{5}) is never satisfied. By assuming that $\sin \phi _{N}=-s$ we find that all solutions will appear at
\begin{equation}
\left| \sin \phi _{n^{\ast }} \right| \geq \left|s\left( \frac{K_{s}^{\text{chain}}}{K}-1\right) \right| .
\label{7}
\end{equation}
The equal sign describes the points where type I solutions touch this
bordering line and is represented by the inferior dotted line in BD.
The solvability region (SR) contains all the synchronized solutions and it can be defined by (\ref{7}). It is worth mentioning  that equation (\ref{7}) shows that the critical synchronization coupling for the ring satisfies the condition $Ks\geq \frac{K_{s}^{\text{chain}}}{2}$.

A closer look at the BD will show that there is always one solution from each bifurcation that is tangent the boundary of the SR (which we call SB) at only one point (this will appear clearly when we discuss figure \ref{figE}): it corresponds to $\sin \phi _{N}=-s$ for type I and $\sin \phi _{n^{\ast }}=s$ for type II solutions. On all of these tangent points, independent of the boundary, a straightforward calculation will show that they satisfy the condition
\begin{equation}
\cos \left( \sum_{n\neq n^{\ast }}^{N-1}\phi _{n}\right) =s\left( \frac{%
K_{s}^{\text{chain}}}{K}-1\right),  \label{8}
\end{equation}
The number of bifurcations depends on the number of solutions of equation (%
\ref{8}) over each boundary of SR. Since the cosine argument may be expanded
as a Laurent series defined by $\phi _{n}=A_{0}+\sum_{m=1}^{\infty }\frac{%
A_{m}^{\left( n\right) }}{K^{m}}$, the decaying behavior as a function of $K$
guarantees a finite number of solutions. From the definition of the phase differences (\ref{3}) it is possible to see that the $A_{0}$ term depends on the size of the system so that the effect of increasing the number of oscillators also increases the number of phase locking solutions above $K_{s}$, as may be observed in figure \ref{figA}.

\begin{figure}[!ht]
\centering\includegraphics[width=1.0\linewidth]{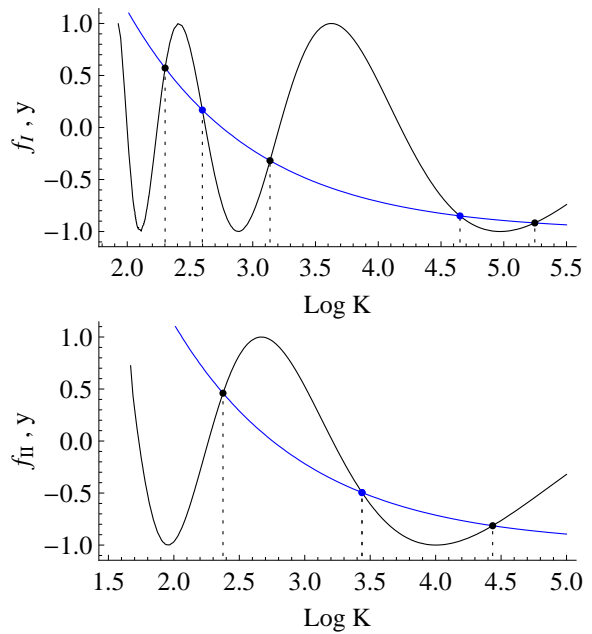}
\caption{(Color online) Graphical representation of the numerical solutions of (\ref{8}) for the $N=10$ realization, with $y = s\left( \frac{%
K_{s}^{\text{chain}}}{K}-1\right)$ (blue curve) and $f_{I,II} = \cos \left( \sum_{n\neq n^{\ast }}^{N-1}\phi _{n}\right)$ (black curve) calculated on the tangent points from type I and II solutions, respectively. Top: $\phi^{\left(I,1\right)}_{n^{\ast }}$ solutions (black dots) at $K_{ \left( m_{1}=0 \right)}=9.9905$, $K_{ \left( m_{1}=1 \right)}=23.0467$, $K_{ \left( m_{1}=2 \right)}=190.016$ and $\phi^{\left(I,2\right)}_{n^{\ast }}$ solutions (blue dots) at $K_{ \left( m_{1}=0 \right)}=13.443$ and $K_{ \left( m_{1}=1 \right)}=104.7171$. Bottom: $\phi^{\left(II,1\right)}_{N}$ solutions (black dots) at $K_{ \left( m_{2}=0 \right)}=10.75473$, $K_{ \left( m_{2}=1 \right)}=84.3981$ and $\phi^{\left(II,2\right)}_{N}$ solutions (blue dot) at $K_{ \left( m_{2}=2 \right)}=31.1029$.}
\label{figB}
\end{figure}

One way of addressing the question of how the multiple solutions are generated is to look at the solutions of $\phi _{n^{\ast }}$ and $\phi _{N}$ on the tangent points. Starting with the type I solutions the condition $\sin{\phi _{N}=-s}$ imposes that $\phi _{N}$ should satisfy the equation
\begin{equation}
\phi^{\left(I\right)}_{N} \left( m_{1}\right)=- s \frac{\left( 4 m_{1}+1 \right) \pi}{2}, \ \ m_{1}=0,1,2,...  \label{1A}
\end{equation}
For each possible value of $m_{1}$ equation (\ref{5}) admits two solutions:
\begin{subequations}
\label{2A}
\begin{eqnarray}
\phi^{\left(I,1\right)}_{n^{\ast }} &=& \arcsin{\left[s\left( \frac{K_{s}^{\text{chain}}}{K}-1\right) \right]}, \label{2Aa} \\
\phi^{\left(I,2\right)}_{n^{\ast }} &=& -\arcsin{\left[s\left( \frac{K_{s}^{\text{chain}}}{K}-1\right) \right]}+\pi. \label{2Ab}
\end{eqnarray}
\end{subequations}
If we fix $\phi_{n^{\ast }}=\phi^{\left(I,1\right)}_{n^{\ast }}$ then the values assumed by $m_{1}$ will provide all the values of $K$ where the solutions with $\cos \phi _{n^{\ast }}>0$ are tangent to the limiting boundary. Similarly when $\phi_{n^{\ast }}=\phi^{\left(I,2\right)}_{n^{\ast }}$ the values of $m_{1}$ give the tangent points for the solutions with $\cos \phi _{n^{\ast }}<0$. Figure \ref{figB} (top) shows these solutions for the $\left\{ \omega \right\}_{N=10}$ realization used previously, when $N = 10$.

The same procedure may be performed for the type II solutions, where the condition $\sin{\phi _{n^{\ast}}=s}$ imposes
\begin{equation}
\phi^{\left(II\right)}_{n^{\ast}} \left( m_{2}\right)=-s \frac{\left( 4 m_{2}-1 \right) \pi}{2}, \ \ m_{2}=0,1,2,...  \label{1A}
\end{equation}
Now for each value of $m_{2}$ we also have two types of solutions for $\phi_{N}$:
\begin{subequations}
\label{3A}
\begin{eqnarray}
\phi^{\left(II,1\right)}_{N} &=& -\arcsin{\left[s\left( \frac{K_{s}^{\text{chain}}}{K}-1\right) \right]}, \label{3Aa} \\
\phi^{\left(II,2\right)}_{N} &=& \arcsin{\left[s\left( \frac{K_{s}^{\text{chain}}}{K}-1\right) \right]}-\pi. \label{3Ab}
\end{eqnarray}
\end{subequations}
The values of $K$ for the tangent points on this boundary are also shown on figure \ref{figB} (bottom), for the same realization with $N = 10$.

With the description of all multiple phase locking solutions above $K_{s}$ (at least on the tangent points) it is possible to have some insight of their origin. If we go back to equation (\ref{4}) and look at the first term on the left hand side it is possible to visualize that as $K$ increases the argument goes beyond $2 \pi$ but not simply adding to it a multiple of $2 \pi$. In this way it is the presence of this term - connecting the first to the last oscillators of the chain - that generates the multiple solutions. The notable symmetry between the type I and type II tangent point solutions comes from the fact that if we remove the interaction term between oscillators $\theta_{1}$ and $\theta_{N}$ the critical synchronization coupling of the resulting chain is the same as the one obtained from extracting the interaction term connecting oscillators $\theta_{n^{\ast}}$ and $\theta_{n^{\ast}+1}$, even when the phase locked solutions may be different. Nevertheless this symmetry is not perfect because the number of type I and type II solutions are in general not the same.

\section{Stability of solutions and basins of attraction}

To perform a linear stability analysis of the solutions it is necessary to obtain the jacobian matrix, but as the equations of motion are invariant by global phase translations $\theta_{n} \rightarrow \theta_{n} + \Theta$ (for every $n$) the analysis is a little more complicated.

It is necessary to realize that the freedom of gauge reduces the $N$ equations of motion (\ref{1}) to a $N-1$ dimensional system (this is an effect of the constraint $ \sum_{n = 1}^{N}\phi _{n} = 0$). Although a gauge fixing condition would enable us to eliminate the extra degree of freedom and perform the stability analysis on the $\theta_{n}$ variables we believe there is a more tractable way to do this. Instead of considering the equations of motion as presented in (\ref{1}) we write the equations on the $\phi_{n}$ variables and eliminate the degree of freedom replacing $\phi_{N}$ by $- \sum_{n = 1}^{N-1}\phi _{n}$. As only the equations for $ \dot{\phi}_{1}$ and $\dot{\phi}_{N-1}$ are dependent on $\phi_{N}$, we have
\begin{subequations}
\label{4A}
\begin{eqnarray}
&& \hspace{-0.7cm} \dot{\phi}_{1} = \omega_{1} - \omega_{2} \nonumber \\
&& \hspace{-0.4cm} -K \left[\sin{\left(\sum_{n = 1}^{N-1}\phi _{n}\right)} +2\sin{\phi_{1}}-\sin{\phi_{2}} \right], \label{4Aa} \\
&& \hspace{-0.7cm} \dot{\phi}_{N-1} = \omega_{N-1} - \omega_{N} \nonumber \\
&& \hspace{-0.4cm} + K \left[ \sin{\left(\sum_{n = 1}^{N-1}\phi _{n}\right)}+ \sin{\phi_{N-2}} -2\sin{\phi_{N-1}}  \right].  \label{4Ac}
\end{eqnarray}
The rest of the equations for $n=2,...,N-2$ do not depend on $\phi_{N}$ and are expressed in the usual form
\begin{eqnarray}
\dot{\phi}_{n} &=&  \omega_{n} - \omega_{n+1}  \nonumber \\
 &+& K \left( \sin{\phi_{n-1}} -2\sin{\phi_{n}} +\sin{\phi_{n+1}}\right).  \label{4Ab}
\end{eqnarray}
\end{subequations}

Due to the structure of the equations, the elements of the jacobian matrix may be written as $J_{n,m} = \partial_{\phi_{m}} \dot{\phi_{n}}$. Since most of the elements are null let us focus on the nonzero ones: the first and last lines are complete with the elements given by
\begin{subequations}
\label{5A}
\begin{eqnarray}
&& \hspace{-0.9cm} J_{1,1} = -K\left[2\cos{\phi_{1}}+\cos{\left(\sum_{j = 1}^{N-1}\phi _{j}\right)} \right], \label{5Ab} \\
&& \hspace{-0.9cm} J_{1,2} = K\left[\cos{\phi_{1}}-\cos{\left(\sum_{j = 1}^{N-1}\phi _{j}\right)} \right],\label{5Ac} \\
&& \hspace{-0.9cm} J_{N-1,N-2} =  K\left[\cos{\phi_{N-2}}-\cos{\left(\sum_{j = 1}^{N-1}\phi _{j}\right)} \right],\label{5Ad} \\
&& \hspace{-0.9cm} J_{N-1,N-1} = -K\left[2\cos{\phi_{N-1}}+\cos{\left(\sum_{j = 1}^{N-1}\phi _{j}\right)} \right], \label{5Ae}
\end{eqnarray}
and for $n=1,...,N-3$
\begin{equation}
J_{N-1,n} = J_{1,n+2} = - k\cos{\left(\sum_{j = 1}^{N-1}\phi _{j}\right)}; \label{5Aa}
\end{equation}
all other lines from $n=2,...,N-2$ have nonzero elements only on the diagonal and first neighbors, namely
\begin{equation}
J_{n,m} = K \cos{\phi_{m}} \left( \delta_{m,n-1} -2\delta_{m,n} + \delta_{m,n+1}\right). \label{5Af}
\end{equation}
\end{subequations}

\begin{figure}[!ht]
\centering\includegraphics[width=1.0\linewidth]{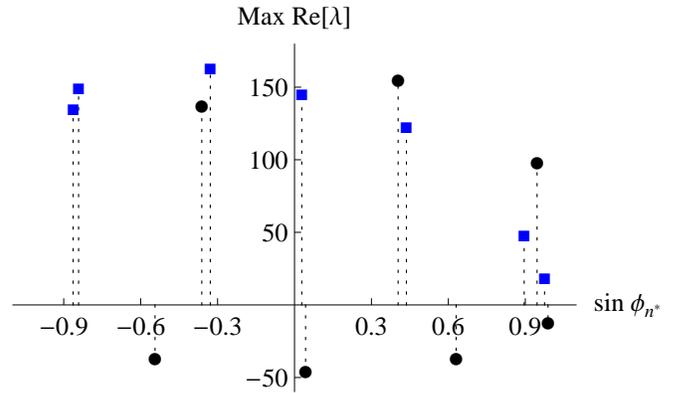}
\caption{(Color online) Largest jacobian eigenvalue for the $N=10$ case with $K=120$. Black dots represent phase locking solutions with $\cos{\phi_{n^{\ast}}}>0$ while blue squares represent phase locking solutions with $\cos{\phi_{n^{\ast}}}<0$. }
\label{figC}
\end{figure}

\begin{figure}[!ht]
\centering\includegraphics[width=1.0\linewidth]{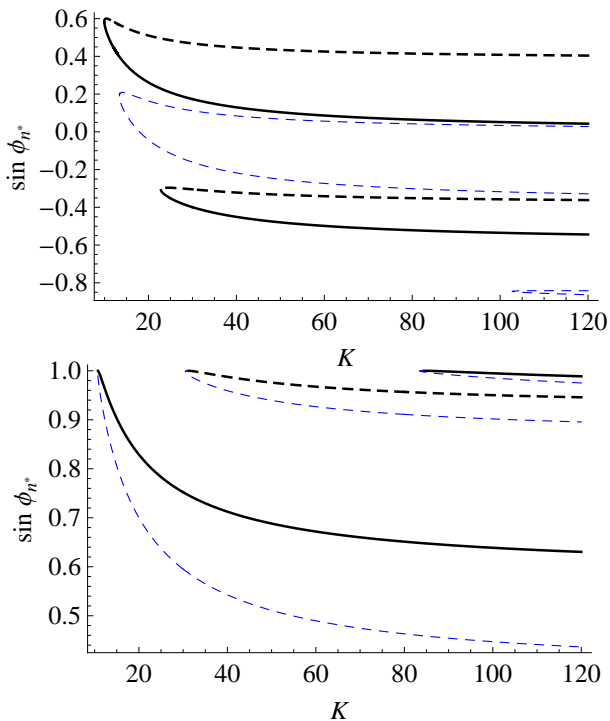}
\caption{(Color online) Bifurcation diagram for $N = 10$ case with explicit stability. Continuous (dashed) curves represent stable (unstable) solutions. Top: type I solutions. Bottom: type II solutions.}
\label{figD}
\end{figure}

\begin{figure}[!ht]
\centering\includegraphics[width=1.0\linewidth]{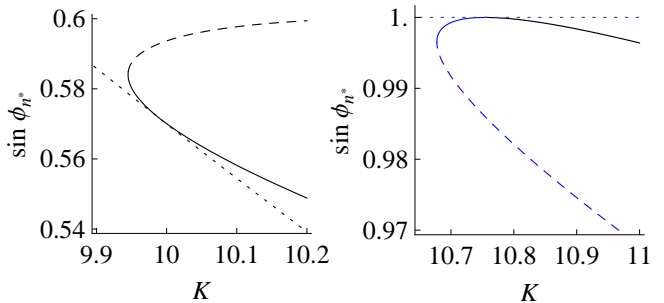}
\caption{(Color online) Zoom of the first two bifurcations present on the BD of the $N = 10$  case. Left: Zoom of the $K_{s}$ bifurcation from figure \ref{figD} (top) with a tangent point from the stable solution touching the SR boundary at $\sin{\phi _{N}}=-s$. Right: first bifurcation from type II solutions from figure \ref{figD} (bottom) showing that the stable solution is actually born with $\cos{\phi_{n^{\ast}}}<0$ and changes sign when it touches the SR curve $\sin \protect\phi _{n^{\ast }}=s$.}
\label{figE}
\end{figure}

The linear stability of the fixed points is determined by the structure of the jacobian eigenvalues: a solution will be stable only if the real part of all eigenvalues are negative. Since a general system with $N$ oscillators has an $\left(N-1\right) \times \left(N-1\right)$ jacobian matrix, the eigenvalue equation is too complicated to be treated analytically. In this way we performed a numerical approach that consists in fixing a value for the coupling $K$ and replacing the phase locked solutions $\phi_{n}$ on the jacobian matrix elements by the numerical calculation of the eigenvalues $\lambda_{n}$, for $n=1,...,N-1$. Figure \ref{figC} shows the real part of the largest eigenvalue for all fixed points for the case $N=10$ with $K=120$ (the fixed points may be inferred either from figure \ref{figA} or figure \ref{figD}). Repeating the procedure for other values of $K$ on the range $\left[K_{s},120 \right]$ we were able to determine the stability of all solutions, as shown on the BD from figure \ref{figD}. Our results confirm that the stability of a branch is not altered by variations of $K$, therefore the results obtained from figure \ref{figC} may be extrapolated to the whole BD.

Let us focus on the $N=10$ case. If we look at the type I solutions (boundary of the SR with $\sin{\phi_{n^{\ast}}} \neq s$), as shown on top of figure \ref{figD}, we observe that if $\cos{\phi_{n^{\ast}}}>0$, two branches are born on a saddle node bifurcation and the stable branch is tangent to the SB curve, $\sin{\phi_{N}}=-s$ (details are shown in Fig.5 left). On the other hand both branches with $\cos{\phi_{n^{\ast}}}<0$ born at the same point, are unstable with at least one positive eigenvalue. If we take a look at the bottom of figure \ref{figD} it is possible to see that the type II solutions (close to $\sin{\phi_{n^{\ast}}} = s$) also present bifurcations that generate two unstable solutions. The stable solutions also have a point tangent to a SR curve ($\sin{\phi_{n^{\ast}}}=s$), but since $\sin{\phi_{n^{\ast}}}=s$, $\cos{\phi_{n^{\ast}}}$ changes sign at the tangent point, different from the type I case, the stable solution starts at the bifurcation with $\cos{\phi_{n^{\ast}}}<0$ and changes sign at the $\sin{\phi_{n^{\ast}}}=s$ line.

Now lets analyze the solutions near the bifurcations. All branches are born at local minima of the function $K\left( \phi_{n^{\ast}}\right)$ defined in the synchronized region ($\dot{\theta}_{n} = \Omega$), where the critical synchronization coupling $K_{s}$ is the absolute minimum. Bearing this in mind, when we take a closer look at bifurcations from each type of solution an interesting feature is observed (figure \ref{5}): $\sin{\phi_{n}}$ is not equal to $\pm 1$ at $K_{s}$ or at any other local minima for any $n$. The explanation for this behavior lies on equation (\ref{5}): in order to have a bifurcation with a sine equal to $\pm1$ it would necessarily be either $\phi_{n^{\ast}}$ or $\phi_{N}$ to present this property, but since $\phi_{N}$ is a nonlinear function of $\phi_{n^{\ast}}$ there are accessible synchronized solutions prior to the appearance of the sine equal to $\pm1$. This result is in agreement with the condition
\begin{equation}
\sum_{n=1}^{N}\frac{\cos \phi _{1}}{\cos \phi _{n}}=0, \label{6A}
\end{equation}
found to be satisfied at the critical coupling $K_s$ by any random distributed natural frequencies on a ring (as shown in \cite{praman}), which makes it impossible to have a bifurcation with a sine equal to $\pm 1$ \cite{01}. Nevertheless there is a sine equal to $\pm1$ near the bifurcation, with the difference becoming smaller as $N\rightarrow \infty$ in agreement with previous literature where this fact has played a crucial role \cite{strog01,hassan01}. On the next section we will show how these deviations and the multiple solutions are generated from a chain of oscillators.

The general picture we obtained for the BD with random distributed frequencies, from both simulation and numerical calculation, may be summarized as follows: for a given N, a configuration $\left\{ \omega \right\} _{N}$ determines a SR for the system in which all solutions come from bifurcations near the solvability boundaries; all stable solutions correspond to branches that touch the solvability boundary; it is possible to extrapolate the definition of the two types of solutions to the bifurcations themselves and label then as $K_{j}^{l}$, where $l=I$ or $II$, for the type I or type II bifurcations and $j=1,2,...$,  gives the order of the minima; odd values of $j$ correspond to saddle node bifurcations while even numbered bifurcations generate two unstable solutions; none of the bifurcations may appear with a sine equal to one because condition (\ref{6A}) must always be satisfied; in addition to the stability analysis our numerical simulations showed that every stable tangent point satisfy the condition $\partial _{K}\cos \left( \sum_{n\neq n^{\ast }}^{N-1}\phi_{n}\right) >0$, although we are not able to explain why it happens at this moment.

With the stability of the solutions fully described we turn our attention to the basin of attraction of the stable solutions. Since a general LCKM is a high dimensional system a usual approach consisting of a graphical analysis becomes extremely difficult either for numerical computation or graphical visualization. To outline these difficulties we consider a statistical approach: given that the fixed points are represented by values of $\sin{\phi_{n^{\ast}}}$ we generate a large sample of random initial conditions for the phases on the interval $\left[-\pi,\pi \right]$, for fixed values of $K$, and estimate the size of the basin of attraction by the probability of the system to reach each of the stable solutions.

\begin{figure}[!ht]
\centering\includegraphics[width=1.0\linewidth]{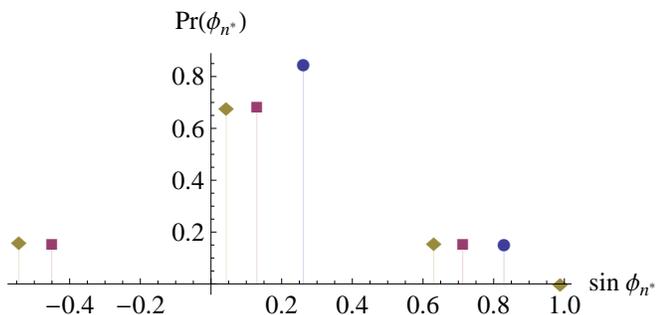}
\caption{(Color online) Estimation of the basin size for each of the stable fixed points within the $N = 10$ case. Blue dots represent $K=20$, lilac squares represent $K=50$ and beige diamonds represent $K=120$. }
\label{figH}
\end{figure}

Starting with the $N=10$ case we could observe that as long as the values assumed by $K$ lie on a region where the number of stable solutions is kept constant, the size of the basins of attraction presents only small statistical fluctuations, which lead us to believe that these variations of $K$ have little effect on the size of a basin. When we vary the coupling constant across regions with increasing number of solutions we observe that when a stable fixed point is created its basin of attraction steals the majority of its size from the closest fixed point (in the $\sin{\phi_{n^{\ast}}}$ space), as it is illustrated on figure \ref{figH} for some values of $K$.

\begin{figure}[!ht]
\centering\includegraphics[width=1.0\linewidth]{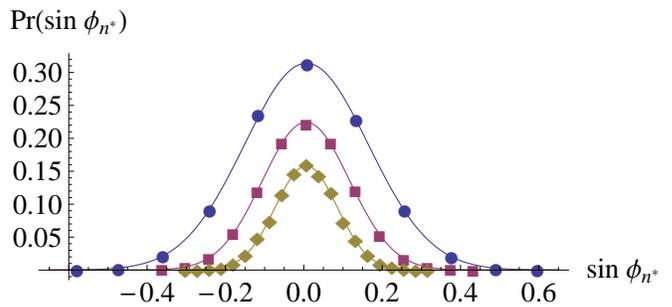}
\caption{(Color online) Probability of the system reaching each of the stable fixed points (for large values of $K$) representing the estimation of the relative basin size. Blue dots represent $N=50$, lilac squares represent $N=100$ and beige diamonds represent $N=200$. }
\label{figI}
\end{figure}

We found numerically that for a given system the size of the basins of attraction is not evenly distributed among all fixed points: the stable solutions with phase locking $\sin{\phi_{n^{\ast}}}$ going to zero as $K \rightarrow \infty$ attract the majority of the initial conditions. Fortunately the behavior of the system for large values of $K$ is easier to analyze: in the limit $K \rightarrow \infty$ we have $\phi_{n}=\phi$, for $n=1,...,N-1$, with $\phi$ given by the solutions of
\begin{equation}
\sin{\left[\left(N-1 \right) \phi\right]}=-\sin{\phi}, \label{7A}
\end{equation}
coming in two types,
\begin{subequations}
\label{8A}
\begin{eqnarray}
\phi_{I} &=& \frac{2 \pi n_{1}}{N},\ \ \ \ \ n_{1}=0,\pm1,\pm 2,..., \label{8Aa} \\
\phi_{II} &=& \frac{\left(2n_{2}+1 \right)\pi }{N-2}, \ \ n_{2}=0,\pm1,\pm 2,... . \label{8Ab}
\end{eqnarray}
\end{subequations}
In this regime the solutions depend only on the system size $N$ and the values assumed by the probability of reaching a stable fixed point are within the curve defined by $f\left( \sin{\phi_{n^{\ast}}} \right) \sim \exp{\left(-a \sin ^{2}\phi _{n^{\ast }} \right)}$, as shown on figure \ref{figI}.

\section{Connection to the chain}

In the previous sections we described the synchronized region of a ring of oscillators and showed how some of the properties of the chain are still present in this topology. Now we focus our attention to understand the appearance of multiple solutions starting from the solution of a chain.

To construct the chain we remove the link between oscillators $1$ and $N$ in the ring to end up with a chain of oscillators where the critical synchronization is defined by $\sin \phi _{n^{\ast }}=s\frac{K_{s}^{\text{chain}}}{K}$ (note that the $\sin \phi _{N}$ term does not enter in the dynamics). A saddle node bifurcation appears at $K=K_{s}^{\text{chain}}$ and naturally we have $\sin \phi _{n^{\ast }}=s$. The effect of closing the chain into a ring by connecting these two oscillators inserts the extra term present on the left hand side of (\ref{5}), which is a nonlinear function of $\phi _{n^{\ast }}$. Instead of a single solution for the fully synchronized state that extends for all $K\geq K_{s}$, this new configuration generates multiple stable solutions born at the local minima of the implicit multivalued function $K\left( \phi _{n^{\ast }}\right) $ that bifurcate into pairs of solutions as $K$ increases.

\begin{figure}
\centering\includegraphics[width=1.0\linewidth]{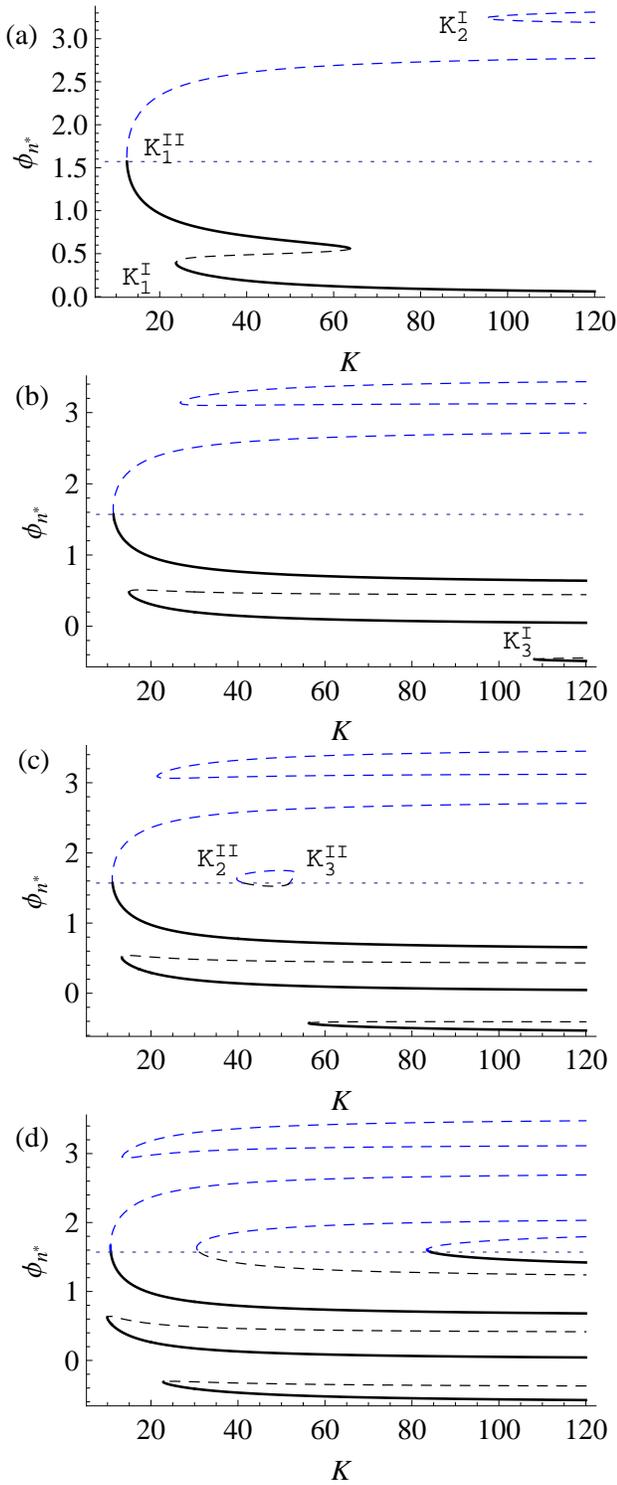}
\caption{(Color online) Bifurcation diagrams $\phi_{n^{\ast }}$ versus $K$ for the $N=10$ case as a function of the
enclosing parameter $\alpha $. (a) Hysteresis figure showing the cusp catastrophe when the stable solution loses stability and disappears; $K_{2}^{I}$ coming from infinity ($\alpha =0.3$). (b) Turning point goes to infinity (bottom) creating two
distinct solutions and $K_{3}^{I}$ comes from infinity ($\alpha =0.5$). (c) Birth of $K_{2}^{II}$ and $K_{3}^{II}$ from the closed circuit ($\alpha =0.7$). (d) Bifurcation diagram of the ring topology showing all solutions on the region ($\alpha =1.0$).}
\label{figF}
\end{figure}

Now we build the ring from the open chain in a controlled way by coupling a continuous parameter $\alpha \in \left[ 0,1\right] $\ to the interaction term $\sin \phi _{N}$ in equation (\ref{2}). It is easy to see for a chain ($\alpha =0$) that $\sin \phi _{n^{\ast }}=s\frac{K_{s}^{\text{chain}}}{K}$ for all $K\geq K_{s}^{\text{chain}}$ \cite{strog01}. The effect of increasing $\alpha $\ deforms the original solution (as shown in figure \ref{figF}) which in turn will generate the others observed for the ring. For small values of $\alpha $ a hysteresis figure appears at a fold bifurcation (fig. \ref{figF}a), creating the $K_{1}^{\text{(I)}}$ bifurcation from the stable branch born at $K_{1}^{\text{(II)}}$, but as $\alpha $ increases the turning point goes to infinity (fig. \ref{figF}b), completely separating the original into two stable solutions. Bifurcations $K_{2}^{\text{(I)}}$\ and $K_{3}^{\text{(I)}}$ do not present such turning points, as they seem to come from $K\rightarrow \infty $\ (figs. \ref{figF}b and \ref{figF}c).

\begin{figure}
\centering\includegraphics[width=1.0\linewidth]{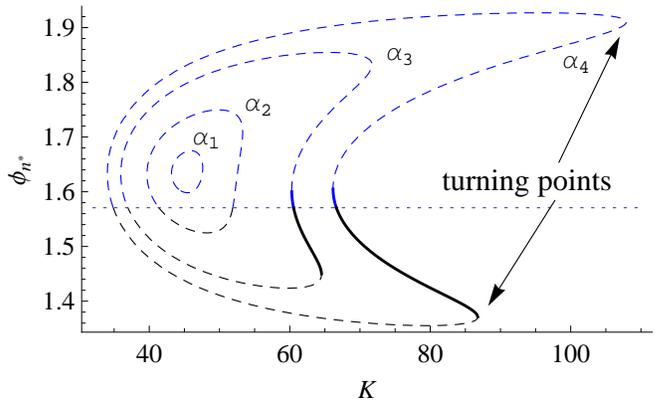}
\caption{(Color online) Closer look at the region of the birth of $K_{2}^{II}$ and $K_{3}^{II}
$ showing solutions for four different values of $\alpha $: $ \alpha _{1}=0.683$, $\alpha _{2}=0.7$, $\alpha _{3}=0.75$
and $\alpha _{4}=0.8$.}
\label{figG}
\end{figure}

Higher values of $\alpha $ give birth to a small closed circuit near $\phi _{n^{\ast }}=\pi /2$ (\ref{figF}c and \ref{figG}), which via a deformation process increases the circuit until the turning points go to $K\rightarrow \infty $, creating $ K_{2}^{\text{(II)}}$\ and $K_{3}^{\text{(II)}}$, each having two solutions.  Figure \ref{figG} shows how the closed circuit of unstable solutions is born ($\alpha =\alpha _{^{1}}$) and as the parameter increases the solution crosses the \ $\phi _{n^{\ast \ }}=\pi /2$ line, creating one stable solution; at this point the bubble starts to deform creating the turning points on both sides of $\pi /2$ and generating four solutions. When $\alpha=1$ the turning points go to infinity and the system presents all the properties of the ring previously described.

\begin{figure}
\centering\includegraphics[width=1.0\linewidth]{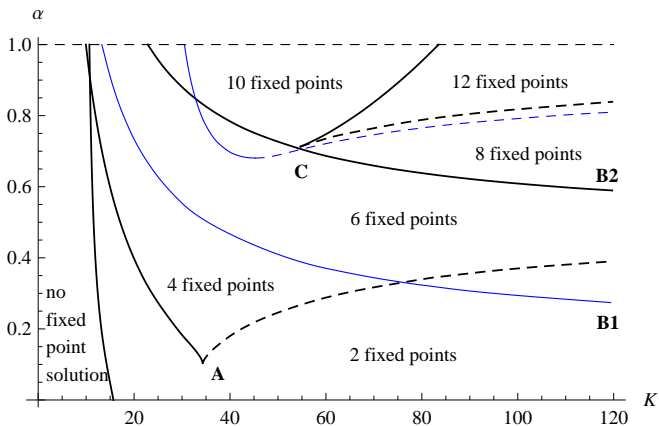}
\caption{(Color online) Stability diagram for the
system. The birth of a pair of solutions is represented by a continuous line
as one goes from lower to higher values of $K$, while the destruction of a
pair of solutions is represented by dotted lines (the definitions are
inverted if one goes from right to left). Thick (black) lines denote the
creation of one stable solution and the thin (blue) ones account for the bifurcations
with no stable equilibria. \textbf{A}: birth of the hysteresis. \textbf{B1}
and \textbf{B2}: bifurcations that come from infinity. \textbf{C}: Birth of
the closed circuit that generates $K_{2}^{\text{(II)}}$\ and $K_{3}^{\text{%
(II)}}$.}
\label{fig05}
\end{figure}

The complete structure of the synchronized state as the system is driven from a chain into a ring may be visualized on the stability diagram shown in figure \ref{fig05}: small values of $\alpha$ change the phase locking solutions $\phi_{n}$ but do not alter the main structure of the chain; for $\alpha \simeq 0.1$ and $K \simeq 35$ (region \textbf{A}) the properties of the ring start to become apparent as the hysteresis appears and the system presents bi-stability on a closed interval of values of $K$; regions \textbf{B1} and \textbf{B2} show the bifurcations $K_{2}^{\text{(I)}}$ and $K_{3}^{\text{(I)}}$ coming from infinity and the disappearance of the hysteresis; region \textbf{C} shows the birth of the closed circuit that creates $K_{2}^{\text{(II)}}$ and $K_{3}^{\text{(II)}}$.

\section{Conclusions}

In conclusion we studied a locally coupled Kuramoto model above the full synchronization transition for a ring of chaotic oscillators. We found a very rich panorama of solutions, although there is only one at the critical value for full synchronization. We were able to determine (analytically) the solvability region (SR) where the solutions exist and to show (numerically) that they all come from minima of the function $K\left( \phi _{n^{\ast \ }}\right)$, with $\phi _{n^{\ast \ }}$ being a specific chosen phase difference. From the observation that every bifurcation has a solution branch with a point tangent to the SR curves we were able to show that the multiplicity of solutions comes as distinct discrete values assumed by $\phi _{n^{\ast \ }}$ and  $\phi _{N}$.

The stability analysis of the solutions showed the existence of saddle node bifurcations (responsible for the creation of the stable solutions) and also bifurcations with only unstable solutions. A statistical approach to estimate the size of the basin of attraction of the stable solutions was performed and we were able to observe that phase locking values of $\sin\phi _{n^{\ast \ }}$ closer to $0$ (for large values of $K$) present the largest basins.

Finally we studied a system with a link coupling strength $\alpha$ varying
from zero (free chain) to one (ring) to investigate the birth of these solutions. We have observed two basic processes responsible for the generation of ring solutions from the open chain: deformations that creates hysteresis for a finite range of $\alpha$; spontaneous creation that either creates solutions coming from infinity or generates closed circuits with four solutions (only one being stable).

F.F.F. thanks the IFT/UNESP for hospitality. P.F.C.T. acknowledges fellowship by CAPES (Brazil).
The authors thank Dr. H.F. El-Nashar for fruitful discussions.


\begin{thebibliography}{10}
\bibitem{01} The presence of symmetries on the $\left\{ \omega \right\} _{N}$
configuration changes this fact, but these results will be published
elsewhere.
\end{thebibliography}
\end{document}